\documentclass[10pt]{article}
\usepackage{graphicx}
\usepackage{amsmath}
\usepackage{amssymb}
\usepackage{caption2}
\begin{document}
\bibliographystyle{prsty}
\begin{center}
{\large {\bf \sc{  Analysis of the fully-heavy pentaquark  states via  the QCD sum rules }}} \\[2mm]
Zhi-Gang  Wang \footnote{E-mail: zgwang@aliyun.com.  }   \\
 Department of Physics, North China Electric Power University, Baoding 071003, P. R. China
\end{center}

\begin{abstract}
In the article, we investigate  the diquark-diquark-antiquark type fully-heavy pentaquark states with the spin-parity $J^P={\frac{1}{2}}^-$ via the QCD sum rules
for the first time, and
 obtain the  masses $M_{cccc\bar{c}}=7.93\pm 0.15\,\rm{GeV}$ and $M_{bbbb\bar{b}}=23.91\pm0.15\,\rm{GeV}$.
  We can search for the fully-heavy  pentaquark  states in the  $J/\psi \Omega_{ccc}$ and  $\Upsilon \Omega_{bbb}$  invariant mass spectrum   in the future.
\end{abstract}



\section{Introduction}
In the quark models, we usually classify the hadrons into the traditional quark-antiquark type mesons, three-quark baryons, and exotic states, such as the
tetraquark (molecular) states,  pentaquark (molecular) states
and hexaquark (molecular) states, etc.
The tetraquark (molecular) states are also referred to as  mesons due to their  integer spins.  A number of exotic states have been observed in recent years, such as the $X(3860)$, $X(3872)$, $Z_c(3885/3900)$, $X(3915)$,  $X(3940)$,
 $Z_c(4020/4025)$, $Z_c(4050)$, $Z_c(4055)$, $Z_c(4100)$,  $X(4160)$, $Z_c(4200)$, $Y(4220)$, $Y(4260)$, $Z_c(4250)$, $Y(4320)$, $Y(4360)$, $Y(4390)$, $Z_c(4430)$,
 $Z_c(4600)$, $Y(4660)$, etc \cite{SLOlsen-review-1708,CPShen-review-1907}.
The most illusive meson $X(3872)$  has hidden-charm, but cannot be
assigned to be  any radial or orbital excited state of the
charmonium, and should have more complicated inner quark
structures  than a mere $c\bar{c}$ pair \cite{FKGuo-review-1705}.
The exotic states provide us with  a unique subject to investigate
the strong interaction, which governs the dynamics of the quarks
and gluons, and the confinement mechanism. If those exotic $X$,
$Y$, $Z$ states are genuine tetraquark (molecular) states, there
are two heavy valence quarks and two light valence  quarks, we
have to deal with the complex dynamics involving both the heavy
and light degrees of freedom.

In 2017, the LHCb collaboration observed the doubly-charmed baryon state  $\Xi_{cc}^{++}$ \cite{LHCb-Xicc},
which provides us with the crucial experimental input on the strong correlation between the two charm (or two heavy) quarks, and is of great importance
on the  spectroscopy of the fully-heavy baryon states and multiquark  states.
    However, just like in the case of the exotic $X$, $Y$, $Z$ states,
    we also have to deal with the complex dynamics involving both the heavy
and light degrees of freedom.  Moreover,  only the
    doubly-charmed baryon state
  $\Xi_{cc}^{++}$ is observed up to now, other doubly-heavy baryon states and triply-heavy baryon states still escape from experimental detecting.

In 2020, the LHCb collaboration  observed a  narrow structure $X(6900)$   and
  a broad structure just above the $J/\psi J/\psi$ threshold in the $J/\psi J/\psi$   invariant mass spectrum \cite{LHCb-cccc-2006},  such structures  are
 the first fully-heavy exotic multiquark candidates claimed  experimentally up to now. It is  a very important step in investigations   of the heavy hadrons,
  and   provides very  important experimental constraints on the theoretical models. The fully-heavy tetraquark candidate $X(6900)$
   was observed before observation of the fully-heavy baryon state $\Omega_{ccc}$.

  The attractive interaction induced by  one-gluon exchange  favors  formation  of  the diquarks in  color antitriplet, which are the basic building blocks of the  diquark-antidiquark type tetraquark
  states \cite{RFLebed-review-1610,AEsposito-review-1611}. Fermi-Dirac
 statistics requires  $\varepsilon^{abc} Q^{T}_b C\Gamma Q_c
  =\varepsilon^{abc} Q^{T}_b \left[C\Gamma\right]^T Q_c$,
  $\left[C\Gamma\right]^T=-C\Gamma$ for $\Gamma=\gamma_5$, $1$,
  $\gamma_\mu\gamma_5$, $\left[C\Gamma\right]^T=C\Gamma$ for
  $\Gamma=\gamma_\mu$, $\sigma_{\mu\nu}$, only the   axialvector  diquarks $\varepsilon^{abc} Q^{T}_b C\gamma_\mu Q_c$ and tensor diquarks
  $\varepsilon^{abc} Q^{T}_b C\sigma_{\mu\nu }Q_c$ can exist, where the $a$, $b$ and $c$ are color indexes.
  Under parity transform $\widehat{P}$,
the diquarks $\varepsilon^{abc}Q^T_bC\gamma_\mu Q_c$ and $\varepsilon^{abc}Q^{T}_bC\sigma_{\mu\nu} Q_c$ have the
properties $\varepsilon^{abc} Q^T_bC\gamma_\mu Q_c
\widehat{P}^{-1}=-\varepsilon^{abc} Q^T_bC\gamma^\mu Q_c$ and $\widehat{P}\varepsilon^{abc}Q^{T}_bC\sigma_{\mu\nu} Q_c\widehat{P}^{-1}=-\varepsilon^{abc}Q^{T}_bC\sigma^{\mu\nu} Q_c$, respectively, so we
usually  refer to them as the axialvector  and tensor diquarks, respectively. The diquarks $\varepsilon^{abc}Q^{T}_bC\sigma_{\mu\nu} Q_c$ have both the $J^P=1^+$ and $1^-$ components, for the $J^P=1^-$ component, there exists an implicit P-wave which is embodied in the negative parity, the axialvector diquarks $\varepsilon^{abc} Q^{T}_b C\gamma_{\mu} Q_c$ are more stable than the tensor diquarks $\varepsilon^{abc} Q^{T}_b C\sigma_{\mu\nu} Q_c$ due to the additional energy excited by the P-wave,
we usually  take the axialvector diquarks $\varepsilon^{abc}  Q^T_bC\gamma_\mu Q_c$ as the basic building blocks to investigate
 doubly-heavy  baryon states, tetraquark states and pentaquark states
 \cite{WZG-AAPPS,WZG-QQ-EPJC-1,WZG-QQ-EPJC-2,WZG-QQ-EPJC-3}.

We can  introduce an explicit P-wave to construct the doubly-heavy diquarks $\varepsilon^{abc} Q^{T}_b
C\gamma_5\stackrel{\leftrightarrow}{\partial}_\mu Q_c$, $\varepsilon^{abc} Q^{T}_b
C\gamma_5\underline{\gamma_5}\stackrel{\leftrightarrow}{\partial}_\mu Q_c$, $\varepsilon^{abc}
Q^{T}_b
C\gamma_{\alpha}\underline{\gamma_5}\stackrel{\leftrightarrow}{\partial}_\mu
Q_c$, which can exist due  to the Fermi-Dirac statistics, there are two-type P-waves, the explicit P-wave is embodied in the 
derivative $\stackrel{\leftrightarrow}{\partial}_\mu=\stackrel{\rightarrow}{\partial}_\mu-\stackrel{\leftarrow}{\partial}_\mu$, while the implicit P-wave is embodied in the underlined $\underline{\gamma_5}$, as multiplying $\gamma_5$ to the diquarks changes their parity. The  $\varepsilon^{abc} Q^{T}_b
C\gamma_5\stackrel{\leftrightarrow}{\partial}_\mu Q_c$ are P-wave diquarks, while the $\varepsilon^{abc} Q^{T}_b
C\gamma_5\underline{\gamma_5}\stackrel{\leftrightarrow}{\partial}_\mu Q_c$ and $\varepsilon^{abc}
Q^{T}_bC\gamma_{\alpha}\underline{\gamma_5}\stackrel{\leftrightarrow}{\partial}_\mu
Q_c$ are D-wave diquarks. We can also introduce an explicit D-wave
$\stackrel{\leftrightarrow}{\partial}_\mu\stackrel{\leftrightarrow}{\partial}_\nu$
to construct the doubly-heavy diquarks, for example, $\varepsilon^{abc} Q^{T}_b
C\gamma_\alpha\stackrel{\leftrightarrow}{\partial}_\mu
\stackrel{\leftrightarrow}{\partial}_\nu Q_c$.
If we take those P-wave and D-wave diquarks as the basic building blocks to investigate the  doubly-heavy  baryon
states, tetraquark states and pentaquark states, we expect to
obtain larger hadron masses considering the additional
contributions  from the P-waves and D-wave.

In 2015,  the  LHCb collaboration  observed  the pentaquark candidates $P_c(4380)$ and $P_c(4450)$ in the $J/\psi p$ mass spectrum
 in the $\Lambda_b^0\to J/\psi K^- p$ decays \cite{LHCb-4380}.
In 2019, the LHCb collaboration  observed a  narrow pentaquark candidate $P_c(4312)$  and proved that the $P_c(4450)$ consists  of
  two overlapping narrow structures $P_c(4440)$ and $P_c(4457)$ \cite{LHCb-Pc4312}.
   In 2020, the LHCb collaboration observed the first evidence of a hidden-charm pentaquark candidate $P_{cs}(4459)$ with  strangeness
   in the $J/\psi \Lambda$  mass spectrum in the $\Xi_b^- \to J/\psi K^- \Lambda$ decays  \cite{LHCb-Pcs4459-2012}.
   They are all very good candidates for the hidden-charm pentaquark (molecular) states \cite{HXChen-review-1601,AAli-review-1706,YRLiu-review-1903}. In this case,
    we also have to deal with the complex dynamics involving  both the heavy and light degrees of freedom.

Analogously, we expect to observe the fully-heavy pentaquark  candidates $QQQQ\bar{Q}$ in the $J/\psi \Omega_{ccc}$ and $\Upsilon \Omega_{bbb}$ invariant mass spectrum
 at the LHCb,  CEPC, FCC and ILC in the future. Theoretically,
J. R. Zhang explores the $\eta_c \Omega_{ccc}$ and $\eta_b \Omega_{bbb}$ pentaquark molecular states with the QCD sum rules \cite{ZhangJR-QQQQQ}.
H. T. An et al explore the fully-heavy pentaquark states  in  the framework of the modified chromo-magnetic interaction model \cite{An-D-QQQQQ}.

In the present work, we construct the diquark-diquark-antiquark type five-quark currents to interpolate the fully-heavy pentaquark states with the same flavor
via the QCD sum rules, and make predictions to be confronted to the experimental data in the future,
as the QCD sum rules is a powerful theoretical tool in investigating the exotic $X$, $Y$, $Z$ and $P$ states \cite{MNielsen-review-1812}.

The article is arranged as follows:  we derive the QCD sum rules for the masses and pole residues
of the  fully-heavy pentaquark states in section 2; in section 3, we present the numerical results and discussions; section 4 is reserved for our conclusion.

\section{QCD sum rules for  the  fully-heavy pentaquark states  }
Let us  write down  the correlation functions   $\Pi (p)$,
\begin{eqnarray}
\Pi(p)&=&i\int d^4x e^{ip \cdot x} \langle0|T\left\{J(x)\bar{J}(0)\right\}|0\rangle \, ,
\end{eqnarray}
where
\begin{eqnarray}
J(x)&=&\varepsilon^{ajk}\varepsilon^{bmn}\varepsilon^{abc}Q^T_j(x)C\gamma_{\mu}Q_k(x)\,Q^T_m(x)C\gamma_{\nu}Q_n(x)\,\sigma^{\mu\nu}C\bar{Q}^T_c(x)\, ,
\end{eqnarray}
$Q=b$, $c$,  the $a$, $b$, $c$, $j$, $k$, $m$, $n$ are color indexes. We can construct other  five-quark currents with the same flavor
to interpolate the fully-heavy pentaquark states,
for example,
\begin{eqnarray}
\eta(x)&=&\varepsilon^{ajk}\varepsilon^{bmn}\varepsilon^{abc}Q^T_j(x)C\sigma_{\mu\alpha}Q_k(x)\,Q^T_m(x)C\sigma^{\mu}{}_{\beta}Q_n(x)\,\sigma^{\alpha\beta}C\bar{Q}^T_c(x)\, , \nonumber\\
\eta_\mu(x)&=&\varepsilon^{ajk}\varepsilon^{bmn}\varepsilon^{abc}Q^T_j(x)C\sigma_{\mu\alpha}Q_k(x)\,Q^T_m(x)C\gamma^{\alpha}Q_n(x)\,C\bar{Q}^T_c(x)\, , \nonumber\\
\eta_{\mu\nu}(x)&=&\varepsilon^{ajk}\varepsilon^{bmn}\varepsilon^{abc}Q^T_j(x)C\sigma_{\mu\alpha}Q_k(x)\,Q^T_m(x)C\sigma_{\nu\beta}Q_n(x)\,\sigma^{\alpha\beta}C\bar{Q}^T_c(x)\, .
\end{eqnarray}
As the  axialvector diquarks $\varepsilon^{ijk} Q^{T}_j C\gamma_{\mu} Q_k$ are more stable than the tensor diquarks $\varepsilon^{ijk} Q^{T}_j C\sigma_{\mu\nu} Q_k$, we choose
the currents $J(x)$ to investigate the lowest fully-heavy pentaquark states.

 Under parity transform $\widehat{P}$, the currents $J(x)$ have the  property,
\begin{eqnarray}
\widehat{P} J(x)\widehat{P}^{-1}&=&-\gamma^0  J(\tilde{x}) \, ,
\end{eqnarray}
where  $x^\mu=(t,\vec{x})$ and $\tilde{x}^\mu=(t,-\vec{x})$.
 The currents $J(x)$  have negative parity, and couple potentially to the fully-heavy pentaquark states with the spin-parity $J^P={\frac{1}{2}}^-$,
\begin{eqnarray}
\langle 0| J (0)|P^{-}(p)\rangle &=&\lambda_{-} U^{-}(p,s) \, ,
\end{eqnarray}
where the $\lambda_{-}$ are the pole residues and the $U^{-}(p,s)$
are the Dirac spinors. However, they also couple potentially to the
fully-heavy pentaquark states with the spin-parity
$J^P={\frac{1}{2}}^+$, because
  multiplying $i \gamma_{5}$ to the currents  $J(x)$ changes their parity,
\begin{eqnarray}
\langle 0| J (0)|P^{+}(p)\rangle &=&\lambda_{+} i\gamma_5 U^{+}(p,s) \, ,
\end{eqnarray}
again the $\lambda_{+}$ are the pole residues and the $U^{+}(p,s)$
are the Dirac spinors.

 At the hadron  side, we isolate the pole terms of the lowest fully-heavy  pentaquark states with  negative parity and positive parity together, and obtain the
 results:
\begin{eqnarray}\label{CF-Hadron-12}
\Pi(p) & = & \lambda^{2}_{-}  {\!\not\!{p}+ M_{-} \over M_{-}^{2}-p^{2}  }+  \lambda^{2}_{+}  {\!\not\!{p}- M_{+} \over M_{+}^{2}-p^{2}  } +\cdots  \, ,\nonumber\\
&=&\Pi_{1}(p^2)\!\not\!{p}+\Pi_{0}(p^2)\, .
 \end{eqnarray}
Then we obtain the hadronic spectral densities  through
dispersion relation,
\begin{eqnarray}
\frac{{\rm Im}\Pi_1(s)}{\pi}&=& \lambda^{2}_{-} \delta\left(s-M_{-}^2\right)+\lambda^{2}_{+} \delta\left(s-M_{+}^2\right) =\, \rho_{1,H}(s) \, , \\
\frac{{\rm Im}\Pi_0(s)}{\pi}&=&M_{-}\lambda^{2}_{-} \delta\left(s-M_{-}^2\right)-M_{+}\lambda_{+}^{2} \delta\left(s-M_{+}^2\right)
=\rho_{0,H}(s) \, ,
\end{eqnarray}
where we add the subscript $H$ to represent  the hadron side. We
introduce the  weight functions
$\sqrt{s}\exp\left(-\frac{s}{T^2}\right)$ and
$\exp\left(-\frac{s}{T^2}\right)$ to acquire  the QCD sum rules at
the hadron side,
\begin{eqnarray}
\int_{25m_Q^2}^{s_0}ds \left[\sqrt{s}\,\rho_{1,H}(s)+\rho_{0,H}(s)\right]\exp\left( -\frac{s}{T^2}\right)
&=&2M_{-}\lambda^{2}_{-}\exp\left( -\frac{M_{-}^2}{T^2}\right) \, ,
\end{eqnarray}
where the $s_0$ are the continuum threshold parameters, the $T^2$ are the Borel parameters, there are no contaminations from the fully-heavy pentaquark states with
 the positive parity due to the special combination.

\begin{figure}
 \centering
 \includegraphics[totalheight=5.0cm,width=12cm]{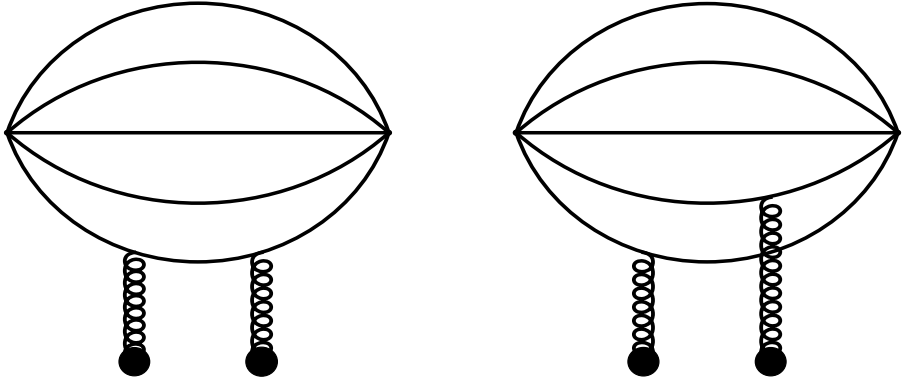}
      \caption{The diagrams contribute  to the gluon condensates. Other
diagrams obtained by interchanging of the $Q$ quark lines  are implied. }\label{Feynman}
\end{figure}

In accomplishing the tedious and terrible operator product expansion, we take account of the gluon condensate and neglect the three-gluon condensate.
After we  obtain  the spectral densities at the quark-gluon level, we match the hadron side with the QCD side of the correlation functions $\Pi(p)$,
again we introduce the weight functions $\sqrt{s}\exp\left(-\frac{s}{T^2}\right)$ and $\exp\left(-\frac{s}{T^2}\right)$  to  obtain  the  QCD sum rules:
\begin{eqnarray}\label{QCDSR}
2M_{-}\lambda_{-}^2\exp\left( -\frac{M_{-}^2}{T^2}\right)&=& \int_{25m_Q^2}^{s_0}ds \int_{16m_Q^2}^{(\sqrt{s}-m_Q)^2}dr \int_{4m_Q^2}^{(\sqrt{r}-2m_Q)^2}dt_1
\int_{4m_Q^2}^{(\sqrt{r}-\sqrt{t_1})^2}dt_2\nonumber\\
&&\rho_{QCD}(s,r,t_1,t_2)\exp\left( -\frac{s}{T^2}\right)\,  ,
\end{eqnarray}
where $\rho_{QCD}(s,r,t_1,t_2)=\sqrt{s}\rho_{1,QCD}+\rho_{0,QCD}$,  $\rho_{0,QCD}=m_Q\widetilde{\rho}_{0,QCD}$,
\begin{eqnarray}\label{QCD-density-1}
\rho_{1,QCD}&=& \frac{1}{2304\pi^8}\frac{\sqrt{\lambda(s,r,m_Q^2)}}{s}\frac{\sqrt{\lambda(r,t_1,t_2)}}{r} \frac{\sqrt{\lambda(t_1,m_Q^2,m_Q^2)}}{t_1}
\frac{\sqrt{\lambda(t_2,m_Q^2,m_Q^2)}}{t_2}\nonumber\\
&&\left( C_{1,8} m_Q^{8}+ C_{1,6} m_Q^{6}+ C_{1,4} m_Q^{4}+ C_{1,2} m_Q^{2}+ C_{1,0}\right)\nonumber\\
&&+\frac{1}{864\pi^6}\langle\frac{\alpha_sGG}{\pi}\rangle\frac{1}{s^2\sqrt{\lambda(s,r,m_Q^2)}^5}\frac{\sqrt{\lambda(r,t_1,t_2)}}{r} \frac{\sqrt{\lambda(t_1,m_Q^2,m_Q^2)}}{t_1}
\frac{\sqrt{\lambda(t_2,m_Q^2,m_Q^2)}}{t_2}\nonumber\\
&&\left(C_{1,16}^{A} m_Q^{16}+ C_{1,14}^A m_Q^{14}+ C_{1,12}^A m_Q^{12}+ C_{1,10}^A m_Q^{10}+C_{1,8}^A m_Q^{8}+ C_{1,6}^A m_Q^{6}+ C_{1,4}^A m_Q^{4}+ C_{1,2}^A m_Q^{2}\right)\nonumber\\
&&+\frac{1}{13824\pi^6}\langle\frac{\alpha_sGG}{\pi}\rangle\frac{\sqrt{\lambda(s,r,m_Q^2)}}{s}\frac{\sqrt{\lambda(r,t_1,t_2)}}{r} \frac{1}{t_1\sqrt{\lambda(t_1,m_Q^2,m_Q^2)}^3}
\frac{\sqrt{\lambda(t_2,m_Q^2,m_Q^2)}}{t_2}\nonumber\\
&&\left( C_{1,10}^{B} m_Q^{10}+C_{1,8}^{B} m_Q^{8}+ C_{1,6}^{B} m_Q^{6}+ C_{1,4}^{B} m_Q^{4}+ C_{1,2}^{B} m_Q^{2}\right)\, ,
\end{eqnarray}
\begin{eqnarray}\label{QCD-density-0}
\widetilde{\rho}_{0,QCD}&=& \frac{1}{1152\pi^8}\frac{\sqrt{\lambda(s,r,m_Q^2)}}{s}\frac{\sqrt{\lambda(r,t_1,t_2)}}{r} \frac{\sqrt{\lambda(t_1,m_Q^2,m_Q^2)}}{t_1}
\frac{\sqrt{\lambda(t_2,m_Q^2,m_Q^2)}}{t_2}\nonumber\\
&&\left( C_{0,4} m_Q^{4}+ C_{0,2} m_Q^{2}+ C_{0,0}\right)\nonumber\\
&&+\frac{1}{1152\pi^6}\langle\frac{\alpha_sGG}{\pi}\rangle\frac{1}{\sqrt{\lambda(s,r,m_Q^2)}^5}\frac{\sqrt{\lambda(r,t_1,t_2)}}{r} \frac{\sqrt{\lambda(t_1,m_Q^2,m_Q^2)}}{t_1}
\frac{\sqrt{\lambda(t_2,m_Q^2,m_Q^2)}}{t_2}\nonumber\\
&&\left( C_{0,6}^A m_Q^{6}+ C_{0,4}^A m_Q^{4}+ C_{0,2}^A m_Q^{2}+C_{0,0}^A\right)\nonumber\\
&&+\frac{1}{13824\pi^6}\langle\frac{\alpha_sGG}{\pi}\rangle\frac{\sqrt{\lambda(s,r,m_Q^2)}}{s}\frac{\sqrt{\lambda(r,t_1,t_2)}}{r} \frac{1}{t_1\sqrt{\lambda(t_1,m_Q^2,m_Q^2)}^3}
\frac{\sqrt{\lambda(t_2,m_Q^2,m_Q^2)}}{t_2}\nonumber\\
&&\left(  C_{0,6}^{B} m_Q^{6}+ C_{0,4}^{B} m_Q^{4}+ C_{0,2}^{B} m_Q^{2}+C_{0,0}^{B}\right)\, ,
\end{eqnarray}
where
\begin{eqnarray}
 C_{1,8} &=&\frac{24}{{s}{r}}-\frac{16({t_1}+{t_2})}{sr^2}+\frac{4r}{st_1t_2}
+\frac{8}{s}\left(\frac{1}{t_1}+\frac{1}{t_2}\right)-\frac{28}{{s}{r}}\left(\frac{t_1}{t_2}+\frac{t_2}{t_1}\right)
+\frac{16}{sr^2}\left(\frac{t_1^2}{t_2}+\frac{t_2^2}{t_1}\right)\, ,\nonumber
\end{eqnarray}
\begin{eqnarray}
 C_{1,6} &=&-\frac{4}{s}+\frac{6({t_1}+{t_2})}{{s}{r}}-\frac{16{t_1}{t_2}}{sr^2}+\frac{32({t_1}+{t_2})-48r}{r^2}
-\frac{8r}{t_1t_2}-\frac{2r^2}{st_1t_2}  -\left(16+\frac{2r}{s}\right)\left(\frac{1}{t_1}+\frac{1}{t_2}\right)   \nonumber\\
&&+\left(\frac{18}{s}+\frac{56}{r}\right)\left(\frac{t_1}{t_2}+\frac{t_2}{t_1}\right)
-\left(\frac{22}{{s}{r}}+\frac{32}{r^2}\right)\left(\frac{t_1^2}{t_2}+\frac{t_2^2}{t_1}\right)
+\frac{8}{sr^2}\left(\frac{t_1^3}{t_2}+\frac{t_2^3}{t_1}\right)\, , \nonumber
\end{eqnarray}
\begin{eqnarray}
 C_{1,4} &=&-28+\frac{11(t_1+t_2)}{s}-\frac{15r}{s}+\frac{24s^2+12s(t_1+t_2)-7(t_1-t_2)^2}{{s}{r}} +\frac{4sr-2r^2}{t_1t_2}-\frac{2r^3}{st_1t_2}   \nonumber\\
&&+\frac{4\left[t_{1}^3+t_{2}^3-t_1t_2(t_1+t_2)-4s^2(t_1+t_2)+8st_1t_2\right]}{sr^2}+\left(8s-8r-\frac{5r^2}{s}\right)\left(\frac{1}{t_1}+\frac{1}{t_2}\right)     \nonumber\\
&&+\left(6+\frac{12r}{s}-\frac{28s}{r}\right)\left(\frac{t_1}{t_2}+\frac{t_2}{t_1}\right)  +\left(\frac{16s}{r^2}+\frac{20}{r}-\frac{1}{s}\right)\left(\frac{t_1^2}{t_2}+\frac{t_2^2}{t_1}\right)
\nonumber\\
&&-\left(\frac{4}{{s}{r}}+\frac{16}{r^2}\right)\left(\frac{t_1^3}{t_2}+\frac{t_2^3}{t_1}\right) \, , \nonumber
\end{eqnarray}
\begin{eqnarray}
 C_{1,2} &=&8s-6r-3(t_{1}+t_{2})+\frac{7(t_{1}^2+t_{2}^2)+10t_{1}t_{2}-9{r}^2}{2s} +\left(2sr-r^2-\frac{r^3}{s}\right)\left(\frac{1}{t_1}+\frac{1}{t_2}\right)    \nonumber\\
&&+\frac{2\left[-s^2(t_{1}+t_{2})-2st_{1}t_{2}+7s(t_{1}^2+t_{2}^2)+t_1t_2(t_1+t_2)-(t_{1}^3+t_{2}^3)\right]}{s{r}}\nonumber\\
&&+\frac{8\left[-t_{1}^3-t_{2}^3+t_{1}t_{2}(t_{1}+t_{2})-2st_{1}t_{2}\right]}{r^2}+\left(4s-2r-\frac{2r^2}{s}\right)\left(\frac{t_{1}}{t_2}+\frac{t_{2}}{t_1}\right)
\nonumber\\
&&+\left(7+\frac{7r}{s}-\frac{14s}{r}\right)\left(\frac{t_{1}^2}{t_2}
+\frac{t_{2}^2}{t_1}\right)+\left(\frac{8s}{r^2}-\frac{4}{r}-\frac{4}{s}\right)\left(\frac{t_{1}^3}{t_{2}}+\frac{t_{2}^3}{t_{1}}\right)\, ,\nonumber
\end{eqnarray}
\begin{eqnarray}
 C_{1,0} &=&sr+2s(t_{1}+t_{2})+\frac{t_{1}t_{2}(2t_{1}+2t_{2}-3s-3r)}{s}-\frac{2(t_{1}^3+t_{2}^3)}{s}
-\frac{{r}({s}+{r})(r+2{r}_1+2{r}_2)}{2s}\nonumber\\
&&+\frac{7({s}+{r})(t_{1}^2+t_{2}^2)}{2s}+\frac{2t_{1}t_{2}(t_{1}+t_{2})+6st_{1}t_{2}-7s(t_{1}^2+t_{2}^2)-2(t_{1}^3+t_{2}^3)}{r}                                       \nonumber\\
&&+\frac{4s\left[t_{1}^3+t_{2}^3-t_{1}t_{2}(t_{1}+t_{2})\right]}{r^2} \, , \nonumber
\end{eqnarray}
\begin{eqnarray}
 C_{0,4} &=&156-\frac{6r^2}{t_{1}t_{2}}-6\left(\frac{t_{1}}{t_{2}}+\frac{t_{2}}{t_{1}}\right)      +12r\left(\frac{1}{t_{1}}+\frac{1}{t_{2}}\right)\, ,\nonumber
\end{eqnarray}
\begin{eqnarray}
 C_{0,2} &=&75(t_{1}+t_{2})+12r+6r\left(\frac{t_{1}}{t_{2}}+\frac{t_{2}}{t_{1}}\right)
-3r^2\left(\frac{1}{t_{1}}+\frac{1}{t_{2}}\right)-3\left(\frac{t_{1}^2}{t_{2}}+\frac{t_{2}^2}{t_{1}}\right)   \, , \nonumber
\end{eqnarray}
\begin{eqnarray}
 C_{0,0} &=&3r(t_{1}+t_{2})+39t_{1}t_{2}-\frac{3}{2}\left(t_{1}^2+t_{2}^2+r^2\right)\, ,\nonumber
\end{eqnarray}
$\lambda(a,b,c)=a^2+b^2+c^2-2ab-2bc-2ac$, the lengthy expressions of the coefficients $C^{A/B}_{1/0,k}$ with $k=16$, $14$, $\cdots$ are neglected for simplicity, the interested readers can get them in the Fortran form or
mathematica form via contact me via E-mail.
 The superscripts $A$ and $B$ correspond to the Feynman diagrams in which the two gluons
 forming the gluon condensate are emitted from one quark line
 (the first diagram in Fig.\ref{Feynman})
 and two quark lines  (the second diagram in Fig.\ref{Feynman}),
 respectively.
From the QCD spectral densities shown in Eqs.\eqref{QCD-density-1}-\eqref{QCD-density-0},
we can see that there are end-point divergences
 $\frac{1}{\sqrt{s-(\sqrt{r}+m_Q)^2}^5}$ and $\frac{1}{\sqrt{t_1-4m_Q^2}^3}$,
 which originate from the first diagram and second diagram in Fig.\ref{Feynman}, respectively.
 In the Appendix, we give some explanations for the origination of
 the end-point divergences.
 In previous works, we observed  that the end-point divergence  $\frac{1}{\sqrt{t_1-4m_Q^2}^3}$ can be regulated by adding a small mass term $\Delta^2$ in
 $\frac{1}{\sqrt{t_1-4m_Q^2+\Delta^2}^3}$  with the value $\Delta^2=(0.2\,\rm{GeV})^2$ \cite{WZG-deta-1,WZG-deta-2,WZG-deta-3}.
  In the present work, we observe that the end-point divergence
 $\frac{1}{\sqrt{s-(\sqrt{r}+m_Q)^2}^5}$ is more severe than the end-point divergence
  $\frac{1}{\sqrt{t_1-4m_Q^2}^3}$,
 we can regulate the divergence by adding a rather large mass term
 $\Delta^{\prime2}$ in $\frac{1}{\sqrt{s-(\sqrt{r}+m_Q)^2+\Delta^{\prime 2}}^5}$ with the value
 $\Delta^{\prime 2}=m_Q^2$.
 Compared to the values of the $(\sqrt{r}+m_Q)^2$, the value
 $\Delta^{\prime 2}=m_Q^2$ is rather  small.
 It is reasonable, as the gluon condensates from
 the first and second Feynman diagrams in Fig.\ref{Feynman} make contributions about the same order.
 In the QCD sum rules for the triply-heavy
 baryon states, the three-gluon condensate makes tiny contributions in the Borel windows, and can be neglected safely \cite{WZG-AAPPS}. Furthermore, in the QCD sum rules for the
 color-singlet-color-singlet type fully-heavy pentaquark states,
 the contributions from the gluon condensate $\langle \frac{\alpha_sGG}{\pi}\rangle$ and
 three-gluon condensate $\langle g_s^3 f^{abc}G^aG^bG^c\rangle$ are very small \cite{ZhangJR-QQQQQ}.
 In the present work, we neglect the contributions from the three-gluon condensate, as it is the vacuum expectation value of the gluon operator of the order $\mathcal{O}(\alpha_s^{\frac{3}{2}})$.
 In the QCD sum rules for the tetraquark (molecular) states and pentaquark (molecular) states, we usually take account of the vacuum condensates which are vacuum
 expectation values of the quark-gluon operators of the
 order $\mathcal{O}(\alpha_s^k)$ with $k\leq 1$
 \cite{WZG-QQQQ-EPJC-1,WZG-QQQQ-EPJC-2,WZG-Hidden-charm-bottom-1,WZG-Hidden-charm-bottom-2,WZG-IJMPA-penta-1,WZG-IJMPA-penta-2}.

We derive   Eq.\eqref{QCDSR} in regard  to  $\frac{1}{T^2}$, then eliminate the
 pole residues $\lambda_{-}$ and obtain the QCD sum rules for
 the masses of the fully-heavy  pentaquark states,
 \begin{eqnarray}
 M^2_{-} &=&-\frac{\frac{d}{d(1/T^2)} \int_{25m_Q^2}^{s_0}ds \int_{16m_Q^2}^{(\sqrt{s}-m_Q)^2}dr \int_{4m_Q^2}^{(\sqrt{r}-2m_Q)^2}dt_1
\int_{4m_Q^2}^{(\sqrt{r}-\sqrt{t_1})^2}dt_2\rho_{QCD}(s,r,t_1,t_2)\exp\left( -\frac{s}{T^2}\right)}{\int_{25m_Q^2}^{s_0}ds \int_{16m_Q^2}^{(\sqrt{s}-m_Q)^2}dr \int_{4m_Q^2}^{(\sqrt{r}-2m_Q)^2}dt_1
\int_{4m_Q^2}^{(\sqrt{r}-\sqrt{t_1})^2}dt_2\rho_{QCD}(s,r,t_1,t_2)\exp\left( -\frac{s}{T^2}\right)}\, .\nonumber\\
\end{eqnarray}

\section{Numerical results and discussions}
We choose  the standard value  of  the  gluon condensate $\langle \frac{\alpha_s
GG}{\pi}\rangle=0.012\pm0.004\,\rm{GeV}^4$
\cite{SVZ79,PRT85,ColangeloReview}, and  take the $\overline{MS}$ masses of the heavy  quarks
$m_{c}(m_c)=(1.275\pm0.025)\,\rm{GeV}$ and $m_{b}(m_b)=(4.18\pm0.03)\,\rm{GeV}$
 from the Particle Data Group \cite{PDG}.
In addition,  we take  account of the energy-scale dependence of the $\overline{MS}$ masses,
 \begin{eqnarray}
 m_Q(\mu)&=&m_Q(m_Q)\left[\frac{\alpha_{s}(\mu)}{\alpha_{s}(m_Q)}\right]^{\frac{12}{33-2n_f}} \, ,\nonumber\\
\alpha_s(\mu)&=&\frac{1}{b_0t}\left[1-\frac{b_1}{b_0^2}\frac{\log t}{t} +\frac{b_1^2(\log^2{t}-\log{t}-1)+b_0b_2}{b_0^4t^2}\right]\, ,
\end{eqnarray}
  where $t=\log \frac{\mu^2}{\Lambda^2}$, $b_0=\frac{33-2n_f}{12\pi}$, $b_1=\frac{153-19n_f}{24\pi^2}$, $b_2=\frac{2857-\frac{5033}{9}n_f+\frac{325}{27}n_f^2}{128\pi^3}$,  $\Lambda=213\,\rm{MeV}$, $296\,\rm{MeV}$  and  $339\,\rm{MeV}$ for the quark flavor numbers  $n_f=5$, $4$ and $3$, respectively  \cite{PDG}.
In the present work, we choose $n_f=4$ and $5$ in the QCD sum rules for the fully-heavy pentaquark states $cccc\bar{c}$ and $bbbb\bar{b}$, respectively, and then
evolve the heavy quark masses  to the  typical energy scales $\mu=m_c(m_c)=1.275\,\rm{GeV}$ and $2.8\,\rm{GeV}$ to extract the  masses of the
 fully-heavy pentaquark states $cccc\bar{c}$ and $bbbb\bar{b}$,
 respectively. Just like in previous works, we add an uncertainties $\delta \mu=\pm 0.1\,\rm{GeV}$ \cite{WZG-DvDvDV}.
 The nonperturbative dynamics are embodied in the running heavy quark masses and gluon condensates.  In Ref.\cite{WZG-AAPPS}, we observe that the best energy scale of the
 QCD spectral density for the triply-bottom baryon state $\Omega_{bbb}$, which has three valence quarks,  is $\mu=2.5\,\rm{GeV}$. In the present case, there are five valence quarks,
 the energy scale of the QCD spectral density for the fully-bottom pentaquark state $bbbb\bar{b}$ should be slightly larger, $\mu>2.5\,\rm{GeV}$, as the pentaquark states are another
 type baryons with the fractional spins. At the typical energy scale $\mu=3.1\,\rm{GeV}$, $m_b(\mu)=4.39\,\rm{GeV}$, which is too small to obtain satisfactory QCD sum rules.
 So we choose $\mu=2.8\pm0.1\,\rm{GeV}$ for the fully-bottom pentaquark state $bbbb\bar{b}$.

We should choose suitable continuum thresholds $s_0$ to  exclude
contaminations from the first radial excited states.  In previous
works, we choose $\sqrt{s_0}= M_{B}+0.50\sim
0.55\pm0.10\,\rm{GeV}$ in the QCD sum rules for the triply-heavy
baryon states $B$ \cite{WZG-AAPPS}, $\sqrt{s_0}=
M_{X}+0.50\pm0.10\,\rm{GeV}$ in the QCD sum rules for the
fully-heavy tetraquark states $X$
\cite{WZG-QQQQ-EPJC-1,WZG-QQQQ-EPJC-2}, $\sqrt{s_0}=
M_{X/Z}+0.55\pm0.10\,\rm{GeV}$ in the QCD sum rules for the
hidden-charm and hidden-bottom tetraquark states $X_Q$ and $Z_Q$
\cite{WZG-Hidden-charm-bottom-1,WZG-Hidden-charm-bottom-2},
$\sqrt{s_0}= M_{P}+0.65\pm0.10\,\rm{GeV}$  in the QCD sum rules
for the hidden-charm pentaquark states $P_c$ and $P_{cs}$
\cite{WZG-IJMPA-penta-1,WZG-IJMPA-penta-2}. The pentaquark states
have
 fractional spins, such as $\frac{1}{2}$, $\frac{3}{2}$, $\frac{5}{2}$, $\cdots$, and they are another type baryon states, in the present work, we choose the
  continuum threshold parameters  $\sqrt{s_0}= M_{P}+0.60\pm 0.10\,\rm{GeV}$ as a rough constraint and vary
 the continuum threshold parameters $s_0$ to search for the best Borel parameters and continuum threshold parameters  to satisfy the two basic criteria of the QCD sum rules
  via trial and error.

 Finally,  we  obtain the optimal continuum threshold parameters $\sqrt{s_0}=8.5\pm 0.1\,\rm{GeV}$ and $24.5\pm 0.1\,\rm{GeV}$ for the fully-heavy pentaquark
  states   $cccc\bar{c}$ and $bbbb\bar{b}$, respectively, and the corresponding Borel parameters are $T^2=4.4-5.4\,\rm{GeV}^2$ and $15.5-18.5\,\rm{GeV}^2$, respectively.
   In the Borel windows, the pole contributions (or the ground state contributions) are about $(44-73)\%$ and $(42-68)\%$ for the $cccc\bar{c}$ and $bbbb\bar{b}$
   pentaquark states,
   respectively,  the pole dominance is satisfied very well. On the other hand,  the contributions of the gluon
   condensate  are about $-7.5\%$ and $< 1\%$ for the $cccc\bar{c}$ and $bbbb\bar{b}$ pentaquark states, respectively,
 the operator product expansion  converges  very well. Now the two basic criteria of the QCD sum rules
 are all satisfied, we expect to make reasonable predictions.

\begin{figure}
 \centering
 \includegraphics[totalheight=6cm,width=7cm]{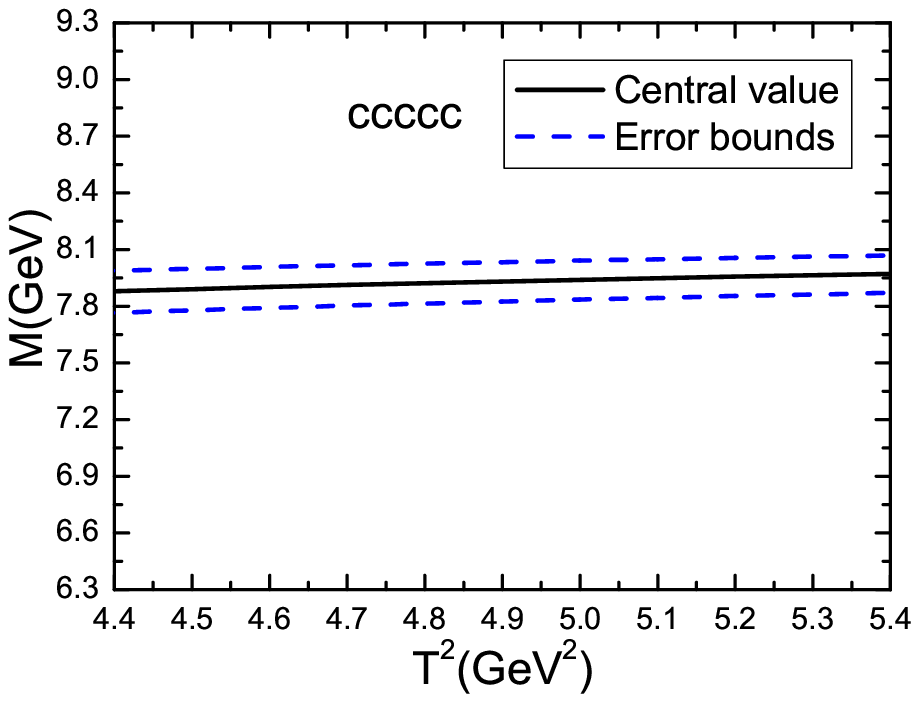}
 \includegraphics[totalheight=6cm,width=7cm]{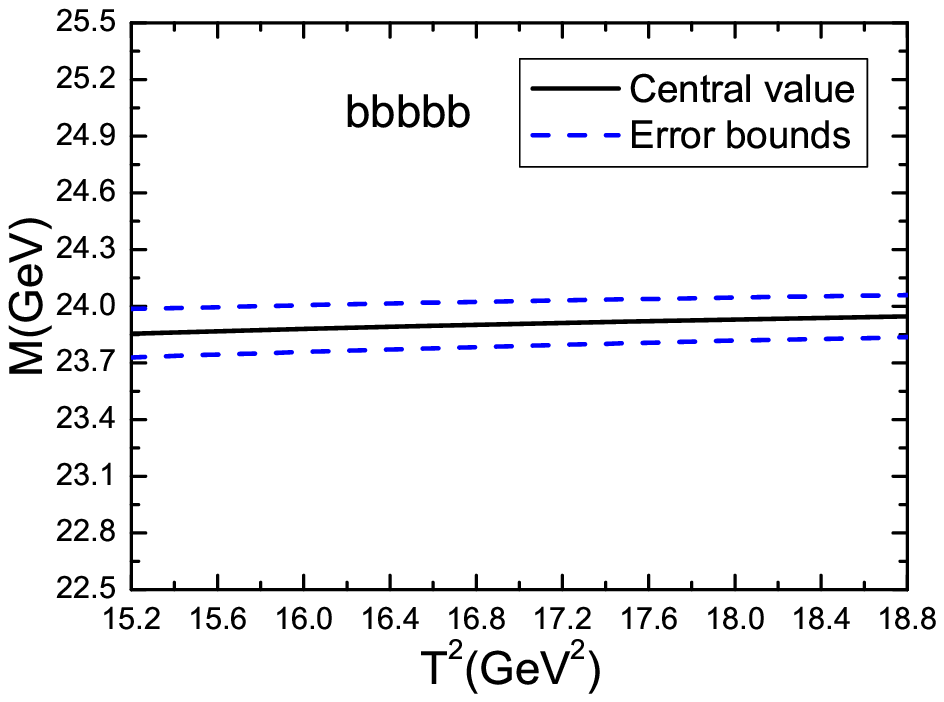}
 \caption{ The masses of the fully-heavy pentaquark states   with variations  of the Borel parameters $T^2$.  }\label{mass-ccccc-bbbbb}
\end{figure}

Then we  take account of all uncertainties of the  parameters,  and obtain the values of the masses and pole residues of the
fully-heavy pentaquark  states,
\begin{eqnarray}
M_{cccc\bar{c}}&=&7.93\pm 0.15\,\rm{GeV}\, ,\nonumber\\
M_{bbbb\bar{b}}&=&23.91\pm0.15\,\rm{GeV}\, ,\nonumber\\
\lambda_{cccc\bar{c}}&=&(0.68\pm 0.21)\times 10^{-1}\,\rm{GeV}^6\, ,\nonumber\\
\lambda_{bbbb\bar{b}}&=&2.43 \pm 0.78\,\rm{GeV}^6\, ,
\end{eqnarray}
which are also shown plainly in Fig.\ref{mass-ccccc-bbbbb}.
 From Fig.\ref{mass-ccccc-bbbbb}, we can see that the predicted masses  are rather stable with variations of the Borel parameters, the uncertainties  come
  from the Borel parameters  are rather small, there appear very flat platforms.

In the QCD sum rules, J. R. Zhang obtains  the predictions of the masses $M_{cccc\bar{c}}=7.38^{+0.20}_{-0.22}\,\rm{GeV}$ and
$M_{bbbb\bar{b}}=21.56^{+0.17}_{-0.15}\,\rm{GeV}$ for the $\Omega_{QQQ}\eta_Q$ type pentaquark molecular states with the spin-parity $J^P={\frac{3}{2}}^-$ \cite{ZhangJR-QQQQQ}.
In the modified chromo-magnetic interaction model, H. T. An et al obtain the predictions of the masses  $7948.8\,\rm{MeV}$ and $7863.6\,\rm{MeV}$ for the
$cccc\bar{c}$ pentaquark states with the spin-parity $J^P={\frac{1}{2}}^-$ and ${\frac{3}{2}}^-$, respectively, and
$23820.7\,\rm{MeV}$ and $23774.8\,\rm{MeV}$ for the $bbbb\bar{b}$ pentaquark states with the spin-parity $J^P={\frac{1}{2}}^-$ and ${\frac{3}{2}}^-$, respectively \cite{An-D-QQQQQ}.
The predictions in Ref.\cite{ZhangJR-QQQQQ} and Ref.\cite{An-D-QQQQQ} are quite different.
The present predictions $M_{cccc\bar{c}}=7.93\pm 0.15\,\rm{GeV}$ and
$M_{bbbb\bar{b}}=23.91\pm0.15\,\rm{GeV}$ for the pentaquark states with the spin-parity $J^P={\frac{1}{2}}^-$ are compatible with that of Ref.\cite{An-D-QQQQQ} within uncertainties.

The decays of the $P_{QQQQ\bar{Q}}$ pentaquark  states can take place through the fall-apart mechanism,
\begin{eqnarray}
P_{cccc\bar{c}}&\to & J/\psi \,\Omega_{ccc} \, , \nonumber\\
P_{bbbb\bar{b}}&\to & \Upsilon \,\Omega_{bbb} \, ,
\end{eqnarray}
according to the predicted masses from the QCD sum rules
\cite{WZG-AAPPS}, we can search for the $P_{QQQQ\bar{Q}}$
pentaquark states in the $J/\psi \Omega_{ccc}$ and $\Upsilon
\Omega_{bbb}$
 invariant mass spectrum  at the LHCb,   CEPC, FCC and ILC in the future. The  triply-heavy baryon states $\Omega_{ccc}$ and $\Omega_{bbb}$ have not been observed yet,
 and we can search for them in the decay chains, $\Omega_{ccc}\to \Omega_{ccs}\,\pi^+ \to \Omega_{css}\,\pi^+\pi^+\to \Omega_{sss}\,\pi^+\pi^+\pi^+$ and
 $\Omega_{bbb}\to \Omega_{bbs}\,J/\psi\to\Omega_{bss}\,J/\psi J/\psi\to\Omega_{sss}\,J/\psi J/\psi J/\psi$ through the weak decays $c \to s u\bar{d}$
and $b\to c\bar{c}s$ at the quark level. We should bear in mind
that the baryon states $\Omega_{ccs}({\frac{1}{2}}^+)$,
$\Omega_{ccs}({\frac{3}{2}}^+)$, $\Omega_{bbs}({\frac{1}{2}}^+)$,
$\Omega_{bbs}({\frac{3}{2}}^+)$ and
$\Omega_{bss}({\frac{3}{2}}^+)$ have also not been observed yet,
we can search for those baryon states as a byproduct.

\section{Conclusion}
In the present work, we construct the diquark-diquark-antiquark type five-quark  currents with the same flavor to study the
fully-heavy pentaquark states with the spin-parity $J^P={\frac{1}{2}}^-$ via the QCD sum rules. After tedious analytical  and numerical calculations,
 we obtain the  masses and pole residues
$M_{cccc\bar{c}}=7.93\pm 0.15\,\rm{GeV}$, $M_{bbbb\bar{b}}=23.91\pm0.15\,\rm{GeV}$, $\lambda_{cccc\bar{c}}=(0.68\pm 0.21)\times 10^{-1}\,\rm{GeV}^6$,
$\lambda_{bbbb\bar{b}}= 2.43 \pm 0.78\,\rm{GeV}^6$. We can search for the fully-heavy  pentaquark  states in the  $J/\psi \Omega_{ccc}$ and
 $\Upsilon \Omega_{bbb}$  invariant mass spectrum  at the  LHCb,  CEPC, FCC and ILC in the future, and confront the predictions to the experimental data.
And we can take the pole residues as the basic input parameters to explore the strong decays of the fully-heavy pentaquark states with the three-point QCD sum rules.

\section*{Appendix}
Now we give an example to illustrate why the end-point divergences
appear in Eqs.\eqref{QCD-density-1}-\eqref{QCD-density-0}. At the
lowest order, we often encounter the typical integral,
\begin{eqnarray}
I_{11}&=&\int d^4k_1
\frac{1}{k_1^2-m_1^2}\frac{1}{(q-k_1)^2-m_2^2}\, ,
\end{eqnarray}
and calculate it by using the Cutkosky's rules,
\begin{eqnarray}
I_{11}&=& \frac{(-2\pi i)^2}{2\pi
i}\int_{(m_1+m_2)^2}^{\infty}dt\frac{1}{t-q^2}\int d^4k_1k^4k_2
\delta^4(k_1+k_2-q)\delta(k_1^2-m_1^2)\delta(k_2^2-m_2^2)\nonumber\\
&=&\frac{(-2\pi i)^2}{2\pi
i}\int_{(m_1+m_2)^2}^{\infty}dt\frac{1}{t-q^2}\frac{\pi}{2}\frac{\sqrt{\lambda(t,m_1^2,m_2^2)}}{t}\,
,
\end{eqnarray}
which is free of end-point divergence. At the second Feynman
diagram in Fig.\ref{Feynman}, we often encounter the typical
integral,
\begin{eqnarray}
I_{22}&=& \int d^4k_1
\frac{1}{(k_1^2-m_1^2)^2}\frac{1}{((q-k_1)^2-m_2^2)^2}\, ,
\end{eqnarray}
again we calculate it by using the Cutkosky's rules,
\begin{eqnarray}
I_{22}&=&\frac{\partial^2}{\partial A \partial B} \int d^4k_1
\frac{1}{k_1^2-A}\frac{1}{(q-k_1)^2-B}\mid_{A\to m_1^2; B\to m_2^2}\nonumber\\
&=&\frac{\partial^2}{\partial A \partial B} \frac{(-2\pi
i)^2}{2\pi
i}\int_{(\sqrt{A}+\sqrt{B})^2}^{\infty}dt\frac{1}{t-q^2}\int
d^4k_1k^4k_2\delta^4(k_1+k_2-q)\delta(k_1^2-A)\delta(k_2^2-B)\nonumber\\
&=&\frac{\partial^2}{\partial A \partial B}\frac{(-2\pi i)^2}{2\pi
i}\int_{(\sqrt{A}+\sqrt{B})^2}^{\infty}dt\frac{1}{t-q^2}\frac{\pi}{2}\frac{\sqrt{\lambda(t,A,B)}}{t}\nonumber\\
&=&\frac{(-2\pi i)^2}{2\pi
i}\int_{(m_1+m_2)^2}^{\infty}dt\frac{1}{t-q^2}\frac{\pi}{2}
\frac{2(m_1^2+m_2^2-t)}{\sqrt{\lambda(t,m_1^2,m_2^2)}^3}\, .
\end{eqnarray}
In the limit $m_1^2=m_2^2=m_Q^2$, we obtain
\begin{eqnarray}
\int_{(m_1+m_2)^2}^{\infty}dt\frac{1}{t-q^2}\frac{1}{\sqrt{\lambda(t,m_1^2,m_2^2)}^3}&=&
\int_{4m_Q^2}^{\infty}dt\frac{1}{t-q^2}\frac{1}{\sqrt{t(t-4m_Q^2)}^3}\,
,
\end{eqnarray}
divergence at the end-point $t=4m_Q^2$ appears. The end-point
divergence  $\frac{1}{\sqrt{s-(\sqrt{r}+m_Q)^2}^5}$ appears at the
first diagram in Fig.\ref{Feynman}, the calculations are
analogous.

\section*{Acknowledgements}
This  work is supported by National Natural Science Foundation, Grant Number  11775079.

\end{document}